\documentclass[aps,prl,reprint,superscriptaddress,nofootinbib,amsmath,amssymb,longbibliography]{revtex4-2}
\usepackage{adjustbox}
\usepackage[colorlinks]{hyperref} 
\usepackage[normalem]{ulem}

\begin{document}

\title{Thermodynamic nature of upbend resonance and validity of Brink-Axel hypothesis in the low-energy region}

\author{L. Tan Phuc}
\email{letanphuc2@duytan.edu.vn}
\affiliation{Institute of Fundamental and Applied Sciences, Duy Tan University, Ho Chi Minh City 70000, Vietnam}
\affiliation{Faculty of Natural Sciences, Duy Tan University, Da Nang City 50000, Vietnam}
\author{N. Quang Hung}
\email{nguyenquanghung5@duytan.edu.vn}
\affiliation{Institute of Fundamental and Applied Sciences, Duy Tan University, Ho Chi Minh City 70000, Vietnam}
\affiliation{Faculty of Natural Sciences, Duy Tan University, Da Nang City 50000, Vietnam}
\author{N. Ngoc Anh}
\email{anh.nguyenngoc1@phenikaa-uni.edu.vn}
\affiliation{Phenikaa Institute for Advance Study (PIAS), PHENIKAA University, Hanoi, 12116, Vietnam}
\author{N. Dinh Dang}
\email{ndinhdang@gmail.com}
\affiliation{Nuclear Many-Body Theory Laboratory, RIKEN Nishina Center for Accelerator-Based Science, 2-1 Hirosawa, Wako City, 351-0198 Saitama, Japan}

\date{\today}

\begin{abstract}  
The nature of low-energy enhancement in the radiative strength function (RSF), which is known as the upbend resonance (UBR) and has a crucial role in the description of neutron-capture cross section and stellar nucleosynthesis, is still under debate. The present Letter extends the exact thermal pairing plus phonon damping model to explore the microscopic nature of the UBR and its thermodynamic origin over a wide mass range of odd-odd, odd-A, and even-even systems, from $^{44}$Sc to $^{153}$Sm, whose experimental RSFs, including the UBRs, are available. The results of our calculations indicate that the UBR arises from non-collective particle-particle and hole-hole excitations with a strength three times stronger than that of the giant dipole resonance. Moreover, the emergence of the UBR at finite temperature provides additional theoretical evidence for the invalidity of the Brink-Axel hypothesis in the very low $E_\gamma$ region. Last but not least, a global relation between the integrated strength of the RSF in the UBR region to that of the total RSF and the mass number is reported, for the first time, within the present study.

\end{abstract}

\maketitle
The radiative strength function (RSF) or $\gamma$-rays strength function (gSF or $\gamma$SF) quantifies the probability that a compound nucleus decays by emitting $\gamma$-rays \cite{Batholomew1973}. This quantity plays a crucial role in the statistical description of $\gamma$-rays-induced reactions, such as the (n,$\gamma$) one. The RSF serves as an essential input for numerous nuclear reactions and nuclear astrophysics calculations \cite{Koning, Rauscher1, Rauscher2, Arnould}, particularly those employing the Hauser-Feshbach model \cite{HauserFeshbach}. However, limited experimental data necessitate the use of theoretical RSFs \cite{Goriely2019}. To develop a reliable RSF model, a deep understanding of its underlying physical mechanism is of paramount importance.
 \par

Experimental RSFs reveal diverse nuclear structural behaviors, mostly nuclear resonances. The latter include the giant dipole resonance (GDR) at high energy ($E_\gamma \approx 15 - 20$ MeV) \cite{Baldwin1947} and lower-energy resonances such as $M$1 spin-flip resonance \cite{Kopecky1993}, pygmy dipole resonance (PDR) \cite{Nyhus}, scissor resonance (SR) \cite{Voinov2001}, and upbend resonance (UBR) \cite{Voinov2004,Larsen2013}. First observed in $^{56,57}$Fe \cite{Voinov2004} and later identified in several nuclei from $^{43}$Sc \cite{Burger2012} to $^{153}$Sm \cite{Simon2016}, the UBR exhibits an unexpected exponential enhancement in the RSF as $ E_\gamma $ approaches zero. Although it mainly affects the low-energy region (typically at $ E_\gamma \leq 3 $ MeV), the UBR significantly impacts the (n,$\gamma$) cross-sections and astrophysical reaction rates with discrepancies exceeding an order of magnitude in neutron-rich nuclei \cite{Larsen2010,Spyrou2024}. Despite extensive studies during the last two decades, the microscopic origin of the UBR remains an open topic. \par
 
Finding the origin of the UBR is challenging as its existence in the very low $E_\gamma$ region seems impossible to be comprehended in theory. No corresponding strength is expected in the ground-state absorption experiments in even-even nuclei due to the pairing property of the nuclear force \cite{Markova2021,Litvinova2013}. However, experimental data have revealed the presence of the UBR in many even-even nuclei \cite{Voinov2004,Larsen2013}. Theoretical descriptions of the UBR are often approached based on the mean field and shell model (SM) calculations. Within the mean-field approaches, attempts have been made to attribute the origin of the UBR to the electric dipole ($E1$) transitions at finite temperature by using the thermal continuum quasiparticle-random-phase approximation (TCQRPA) \cite{Litvinova2013} and finite-temperature relativistic time blocking approximation (FT-RTBA) \cite{Wibowo}. The obtained results, although showing an enhancement of RSF at low $E_\gamma$, cannot convincingly describe the whole experimental RSF data. Recent advanced calculations using combined many-body methods, such as static-path approximation plus RPA (SPA+RPA), quasiparticle RPA (QRPA) with and without temperature dependence, shell-model Monte Carlo (SMMC) with maximum entropy method (MEM), and self-consistent finite-temperature relativistic QRPA (FT-RQRPA) \cite{Alhassid1,Alhassid2,Alhassid3,Frosini,Kaur}, face similar limitations. For instance, to describe the RSF, the recent FT-RQRPA approach \cite{Kaur} employed two reference parameters, $C_E$ and $C_G$, to obtain reasonable values for the resonance energy $E$ and its width $\Gamma$. In addition, to describe the $M1$ strength function in the low-energy region, $\Gamma_{M1}$ is artificially fixed at 1 MeV. Consequently, the spin-flip $M1$ resonances ($E_\gamma \approx 6-10$ MeV) are very narrow, whereas they are typically much broader, with width of about 4 MeV). Within the SM framework, the $M1$ nature of the UBR has been insistently suggested. However, the origin of these $M1$ transitions is still under debate and falls into two main interpretations. The foremost SM study for Mo isotopes \cite{Schwengner} tied the UBR to the recoupling of spins of protons and neutrons with high-$j$ orbitals, which is similar to the shears mechanism in magnetic rotation \cite{Frauendorf2001}. A later study for Ge isotopes \cite{Frauendorf2022} revealed that the UBR is strongest near closed shells, especially at the almost completely filled 1$g_{9/2}$ orbitals. Alternatively, a competing interpretation suggested that the strong low-energy $M$1 transitions in $^{56,57}$Fe nuclei are generated by the high-$l$ terms in the 0$\hbar\omega$ transitions \cite{Brown}. Other SM-based studies \cite{Schwegner2017,Midtb2018,Larsen2018,Fanto2024} proposed the coexistence of the UBR and the well-known SR. Nevertheless, the connection between the shears-like mechanism, which is supposed to cause the UBR, and the scissors vibration \cite{Iudice1978}, which generates the SR, remains unclear, despite both involving the neutron-proton reorientation. The latest advancement \cite{Chen}, employing an angular-momentum-projected SM, has proposed that the UBR originates from a quasi-free scissors motion occurring only in weakly deformed nuclei, similar to free-rotation of neutrons with respect to protons. This mode is introduced as a new type of collective mode called scissors rotation. However, three limitations still remain in this study, namely i) the UBR has been observed in not only weakly deformed nuclei but also well-deformed ones, such as $^{151,153}$Sm \cite{Simon2016}; ii) only three Nd isotopes evidence its broad validity claim; and iii) theoretical descriptions of the UBR do not closely match experimental data (see Fig. 1 of Ref. \cite{Chen}).
\par

In the RSF study, the Brink-Axel hypothesis (BAH) \cite{BAH}, which generally states that the RSF does not depend on the energies, spins and parities of initial and final states (it depends on $E_\gamma$ only), is also a topic of interest. Although the BAH plays an important role in both fundamental and applied nuclear physics, its validity, despite confirmations from various studies \cite{Oslodata,Markova2021,BAHtestexp1,BAHtestexp2,BAHtestexp3,BAHtestexp4,BAHtestexp5,BAHtesttheor1,BAHtesttheor2,BAHtesttheor3,BAHtesttheor4}, remains uncertain with open questions persisting, especially in the context of nuclear astrophysics
\cite{Nabi,Farooq,Herrera,Sieja2023,Misch,HungPRL,Johnson,Angell,Isaak2013,Isaak2019,Netterdon,Wasilewska,Loens2012,Sieja2018,Isaak2018,Mercenne2024,Gorton2026}. Recent studies highlight deviations exceeding three orders of magnitude related to the use of BAH in calculating the stellar electron capture and $\beta-$decay rates at high temperature and high matter density \cite{Nabi,Farooq}. Given these uncertainties, validating the BAH is crucial for improving the RSF models and their applications. While the BAH seems to be well-established in the GDR region, its validity in lower energy regions ($E_\gamma \leq B_n$), such as the PDR and below, remains debated. Currently, as mentioned, only the SM \cite{Schwengner,Brown,Chen} have successfully explained the UBR formation, whose results do not conflict with the BAH, but several studies have shown that the BAH is hardly valid in such a low-energy region \cite{Sieja2023,Misch,HungPRL}.

One of the effective microscopic approaches to the RSF within the quasiparticle framework has been proposed in Ref. \cite{HungPRL}, where the exact thermal pairing plus phonon damping model (EP+PDM) is employed. This approach has shown a strong temperature dependence in the RSF at low energies, raising an open question regarding the validity of the BAH. The EP+PDM, which accounts for both the temperature-dependent GDR width (via the PDM) and thermal pairing effects (via the EP), has been successfully employed to microscopically describe the whole experimental RSFs of several nuclei \cite{HungPRL,EPPDM1,EPPDM2,Phuc2024,Senapati}. The PDM, which is constructed based on the interaction of nuclear collective vibration (phonon) with the single-particle states through the response of the double-time Green function \cite{PDM1,PDM2}, represents the $\gamma$ transitions as couplings between particle ($p$) and hole ($h$) states at $T \geq 0$. The $pp$ and $hh$ excitations, which are forbidden at $T=0$, only contribute at $T>0$ and thus cause the broadening of the GDR width with increasing $T$. These couplings are also argued to be tantamount to the contribution of thermal shape fluctuations \cite{PDM2}. Notably, the EP+PDM naturally captures part of SR \cite{Phuc2024} and PDR \cite{HungPRL,EPPDM1} without extra parameters. However, this model constrains the entire gamma spectrum to a single oscillation regime of the GDR phonon, resulting in a concentrated distribution of collective transitions around the GDR region. In order to describe the UBR at the very low $E_\gamma$ region, the EP+PDM needs to be extended. The aim of this Letter is to extend the EP+PDM to explore the thermodynamic nature of the UBR as well as to reexamine the validity of the BAH in such a low $E_\gamma$ region. \par

To microscopically describe the UBR within the EP+PDM, we treat the UBR as a collective excitation in the very low $E_\gamma$ region, termed the UBR phonon. This phonon is completely separated from the GDR one and only exists in nuclei that exhibit the UBR. In such nuclei, both the RSF of the GDR phonon in the high-energy region ($f^{\rm GDR}$) and that of the UBR one in the very low-energy region ($ f^{\rm UBR}$) contribute to form the main part of the total RSF. In general, the couplings to $ph$, $pp$, and $hh$ configurations cause the damping of these collective phonons. As a result, each of them acquires a width, which is a sum of quantal and thermal widths. The quantal width is caused by the couplings to $ph$ configurations, which contribute already at $T=$ 0, whereas the couplings to $pp$ and $hh$ configurations emerge only at $T\neq$ 0 due to the distortion of the Fermi-surface. In our model, the GDR and UBR phonons are treated within a unique statistical ensemble, so $f^{\rm GDR}$ and $ f^{\rm UBR}$ share a common temperature variable.	

\begin{figure}
       \includegraphics[scale=0.29]{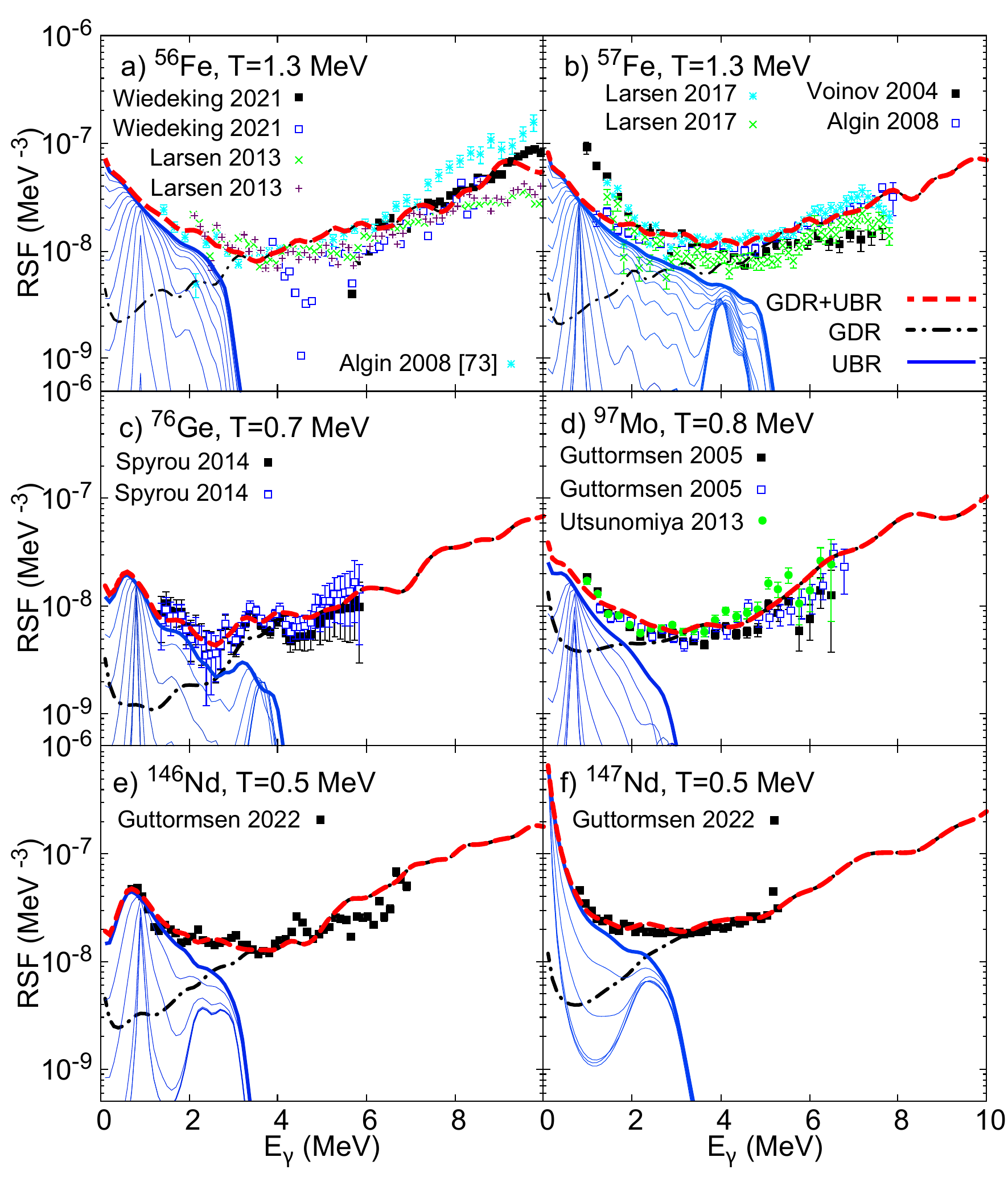}
       \caption{Comparison between the RSFs of $^{56}$Fe (a), $^{57}$Fe (b), $^{76}$Ge (c), $^{97}$Mo (d), $^{146}$Nd (e), and $^{147}$Nd (f) obtained within the extended EP+PDM at different temperatures with experimental data. GDR and UBR denote $f^{\rm GDR}$ and $f^{\rm UBR}$, respectively. Thin lines under the solid curve represent the values of $f^{\rm UBR}$ calculated at different temperatures, from $T=0$ to the $T$ value at which $f^{\rm GDR}$ is defined, with a step of $\delta T=0.1$ MeV. Details are given in the text. Experimental data are from Voinov 2004: \cite{Voinov2004}, Algin 2008: \cite{Algin}, Larsen 2013: \cite{Larsen2013}, Larsen 2017: \cite{Larsen17}, Wiedeking 2021: \cite{Wiedeking21}, Spyrou 2014: \cite{Spyrou14}, Guttormsen 2005: \cite{Mo2005}, Utsunomiya 2013: \cite{Mo2013}, and Guttormsen 2022: \cite{Guttormsen22}.}
        \label{Fig1}
    \end{figure}

Following the formalism, discussed in detail in Refs. \cite{HungPRL,EPPDM1,EPPDM2}, the RSF $f^{R}(E_{\gamma}, T)$ ($R =$ GDR or UBR) is described as 
\begin{equation}
\label{fUB}
f^{R}(E_{\gamma},T)=\frac{\pi}{3\pi^2 \hbar^2 c^2} \frac{\sigma_{R}\gamma^{R}_q(E_{\gamma},T)S^{R}(E_{\gamma},T)}{E_{\gamma}}~, 
\end{equation}
where $E_{R} $ and $\sigma_{R}$ are the resonance energy and cross-section, respectively. The resonance strength function $S^{R}(E_{\gamma},T)$ is given as
\begin{equation}
\label{SUB}
S^{R}(E_{\gamma},T)=\frac{1}{\pi}\frac{\gamma_q^{R}(E_{\gamma},T)}{(E_{\gamma}-E_{R})^2+[\gamma^{R}_q(E_{\gamma},T)]^2}~.
\end{equation} 
The damping of the UBR is microscopically described as follows,
\begin{eqnarray}
\hspace{-5mm} \label{gammaPDMUB}
&&\gamma_q^{\rm UBR}(E_\gamma,T)= \\
&& \pi [F^{\rm UBR}_{ph}]^2\sum_{ph}[u_{ph}^{(+)}]^2(1-n_p-n_h)\delta(E_\gamma-E_p-E_h) ~\nonumber \\
&+&\pi [F^{\rm UBR}_{ss'}]^2\sum_{s>s'}[v_{ss'}^{(-)}]^2(n_s-n_{s'})\delta(E_\gamma-E_s+E_{s'})~, \nonumber
\end{eqnarray}
where $E_{p(h)}$ and $n_{p(h)}$ are quasiparticle energies and
quasiparticle occupation numbers of the $p$($h$) states, respectively, whereas the notation $ss'$ stands for $pp'$ or $hh'$. The combination of Bololyubov coefficients $u^{(+)}_{ph} = u_pv_h + v_pu_h$ and $v^{(-)}_{ss'} = u_su_{s'} - v_sv_{s'}$ are consistently used as in Refs. \cite{EPPDM1,EPPDM2,PDM2,DangPRC12}. The value of $E_{\rm UBR}$ for each nucleus is determined based on the empirical $M1$ RSF, which is evaluated by using the experimental discrete level schemes following the methodology presented in Ref. \cite{Midtb2018}. Specifically, individual $M1$ transitions are identified between specific excited states with well-established spin and parity assignments, and their strengths are extracted from measured initial-state lifetimes and experimental branching ratios. To account for the statistical nature of the electromagnetic decay and to mitigate Porter-Thomas fluctuations, these individual strengths are averaged over narrow energy intervals. Once empirical $E_{\rm UBR}$ values are determined, their global dependence on the mass number $A$ can be obtained via fitting functions. As for $\sigma_{\rm UBR}$, this parameter can be microscopically extracted if the percentage of the total energy-weighted sum rule exhausted by the UBR is known. Alternatively, it can also be determined through a global relation with $A$, similar to $E_{\rm UBR}$. The formation of the UBR can be microscopically described via the radiative width given in Eq. (\ref{gammaPDMUB}). In the latter, the coupling of the UBR phonon to $ph$ configurations correspond to the quantal damping, whereas the coupling to $pp$ and $hh$ ones stands for the thermal damping. We note that the contributions of the quantal and thermal components are maximal at the $ph$ poles ($E_p + E_h$) and $pp$ or $hh$ ones ($E_s - E_{s'}$), respectively. Because the $ph$ poles $E_p + E_h\geq 2\Delta$ where $\Delta$ is the pairing gap at $T=0$, the couplings to $ph$ configurations have no practical contribution at $E_\gamma < 2\Delta$. At the same time, the $pp$ and $hh$ poles $E_s - E_{s'}$ can be very close to zero when $E_s$ is close to $E_{s'}$, and there are many of them. Therefore, at $T\neq$ 0, when the differences $n_s - n_{s'}$ in the numerators of the second term at the RHS of Eq. (\ref{gammaPDMUB}) become finite and increase with $T$, the contribution of the $pp$ and $hh$ configurations to the damping of the UBR becomes paramountly important. This leads to the formation of a ``local” collective oscillation nowhere but in the very low $E_\gamma$ region. In the results discussed below, at first $f^{\rm GDR}$ is calculated by using the global Steinwedel-Jensen parameters for the GDRs, and the temperature is chosen around those predicted by the constant-temperature (CT) model \cite{CT} or those extracted from the experimental discrete level scheme. The temperature-independent parameters $F^{\rm GDR}_{ph}$ and $F^{\rm GDR}_{ss'}$ are determined following the description in Refs. \cite{PDM1,PDM2}, ensuring the microscopic nature of the model. As for the UBR part, a systematic investigation of the UBR across a wide range of nuclei, from light to heavy atomic masses (see, e.g., Fig. \ref{Fig1}), reveals that the values of the corresponding $F^{\rm UBR}_{ph}$ and $F^{\rm UBR}_{ss'}$ strengths are also temperature-independent and consistently three times larger than those corresponding to the GDR. 

\begin{figure}
       \includegraphics[scale=0.25]{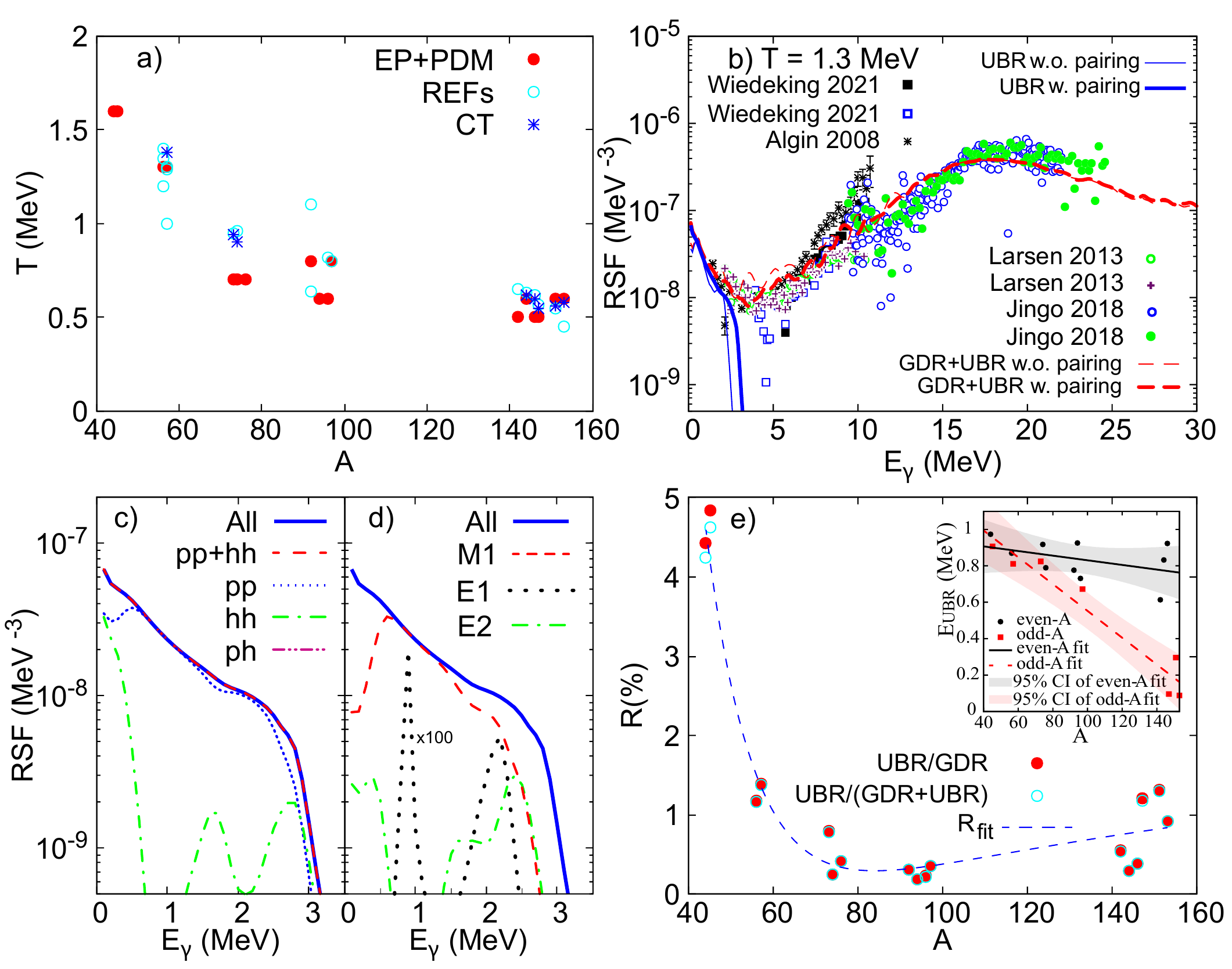}
       \caption{(a): Comparison of nuclear temperature obtained within the extended EP+PDM with those reported by the CT model \cite{CT} and other references \cite{Wiedeking21,Larsen17,Algin,Ge,Tveten,Simon2016} for various UBR nuclei. (b): The RSFs of $^{56}$Fe with and without exact pairing (EP). Experimental data are from Algin 2008: \cite{Algin}, Larsen 2013: \cite{Larsen2013}, Wiedeking 2021: \cite{Wiedeking21}, and Jingo 2018: \cite{Jingo18}. (c): Contribution of $ph$, $pp$, and $hh$ configurations to the UBR in the low $E_\gamma$ of (b). (d): Multipole contribution of $M$1 excitation to the UBR in (c), obtained within the extended EP+PDM. The E1 contribution is multiplied by a factor of 100 for clarity, owing its significantly small magnitude. (e): Weighted $f^{\rm UBR}/f^{\rm GDR}$ and $f^{\rm UBR}/f^{\rm GDR+UBR}$ ratios obtained within the extended EP+PDM for various UBR nuclei. Inset in (e): linear relations between $E_{\rm UBR}$ and $A$ for odd- and even-$A$ systems (CI: confident interval).
        \label{Fig2}}
    \end{figure}

Figure \ref{Fig1} compares the RSFs calculated using the extended EP+PDM at defined temperatures for six nuclei, including odd-A ($^{57}$Fe, $^{97}$Mo, and $^{147}$Nd) and even-even ($^{56}$Fe, $^{76}$Ge, and $^{146}$Nd), whose experimental data were obtained using the Oslo method  
\cite{Oslodata,Wiedeking21,Larsen2013,Algin,Voinov2004,Larsen17,Spyrou14,Mo2005,Mo2013,Guttormsen22}. The input single-particle spectra are calculated from the axially deformed Woods-Saxon potential \cite{WS}, which provides 126 doubly degenerate (deformed) levels (corresponding to approximately eight major shells) for both protons and neutrons. The exact thermal pairing (EP) calculations are performed within a truncated model space around the Fermi surface where pairing correlations play the dominant role, as discussed in Refs. \cite{HungPRL,EPPDM1}. In addition, the $f^{\rm UBR}$ calculations are only applied to the energy region where the UBR exits. It is clear to see in Fig. \ref{Fig1} that the total RSFs calculated within the extended EP+PDM (red dashed lines) are in excellent agreement with experimental data in the whole energy range. For $^{57}$Fe, our theoretical prediction agrees with the latest experimental data in Ref. \cite{Larsen2013}. The fact that the temperatures at which the total RSFs are calculated, i.e. $T=1.3, 0.7, 0.8$ and $0.5$ MeV, respectively for $^{56,57}$Fe, $^{76}$Ge, $^{97}$Mo, and $^{146,147}$Nd, are close to those predicted by the CT model \cite{CT} and other references (see Fig. \ref{Fig2}(a)), guarantees the microscopic nature of the model proposed. At those defined temperatures, $f^{\rm GDR}$ (black dash-dotted lines) aligns with experimental data at high energies but remains significantly lower at low energies. The flat contribution of $f^{\rm GDR}$ in the low-energy region is also in line with that reported in Ref. \cite{Sieja2017}. In contrast, $f^{\rm UBR}$ (blue thick lines), calculated at the same temperatures with $f^{\rm GDR}$, can reproduce well the very low-energy data. One can also see in Fig. \ref{Fig1} the evolution of $f^{\rm UBR}$ with $T$ (blue thin lines), in which $f^{\rm UBR}$ is absent at $T=0$ and gradually increases with increasing $T$ to match the low-energy data. This implies the purely thermodynamic nature of this resonance.

To further investigate the nature of the UBR, Fig. \ref{Fig2}(b) shows the RSFs of $^{56}$Fe along with the separated contributions from the $ph$, $pp$, $hh$, and $pp+hh$ configurations [Fig. \ref{Fig2}(c)], and those among them, which form the $M$1 excitation [Fig. \ref{Fig2}(d)] to the UBR at $T=1.3$ MeV. As seen in Fig. \ref{Fig2}(b), while the total RSF obtained with exact pairing (EP) nicely agrees with the experimental data in the entire energy region \cite{Wiedeking21,Jingo18}, that obtained without EP overestimates the data in the intermediate energy region, namely $E_\gamma \approx 3-10$ MeV. This highlights the crucial role of the EP in describing the RSF below $B_n$ (11.2 MeV for $^{56}$Fe). Meanwhile, the $f^{\rm UBR}$ values obtained with and without EP are nearly the same in the UBR region ($E_\gamma\le 3$ MeV), thus the EP has almost negligible contribution to the UBR. Instead, the UBR is mainly contributed by the $pp$ and $hh$ excitations, in which the contribution of $pp$ configurations is strongly dominant, as revealed in Fig. \ref{Fig2}(c). The contribution of $hh$ configurations, although small, is still important and cannot be ignored. This observation reveals the single-particle structure, where the $h$ states lie in the bound-state region (sparse in light nuclei), whereas the $p$ states are easily excited to the continuum as temperature increases. We also note that the energies of such $pp$ and $hh$ excitations are often very low, thus they can be viewed as the nearly free motion of non-collective nucleons around a collective oscillation center, similar to the mechanism proposed in Ref. \cite{Chen}. Another unresolved question concerning the UBR is whether its nature is electric dipole, magnetic dipole, or a mixing of both. Fig. \ref{Fig2}(d) demonstrates that the UBR obtained within the extended EP+PDM is of dipole nature and mainly contributed by $M1$ excitations, while the $E1$ contribution is negligible and the $E2$ strength is about one or two orders of magnitude smaller than the $M1$ one, particularly at 0.5 MeV $< E_\gamma <$ 1.5 MeV. Because of the deviation of $M$1 and UBR strengths, the UBR may contain a mixture of higher multipole excitations, namely the integrated strength ratio $\int f^{UBR}_{M1}dE_\gamma /\int f^{UBR}_{\rm All}dE_\gamma \simeq$ achieves about $60\%$. This result is in line with the latest large-scale shell model (LSSM) calculations, which indicate that the $M1$ transitions are not merely the source of the enhanced gamma strength below 3 MeV \cite{Gorton2026}.

Compared to existing models, the present approach offers two key advantages: i) it is free from spurious states that frequently appear in the RPA-based calculations and can be mistakenly associated with the UBR region, and ii) it goes beyond the RPA, which cannot describe the resonance width and its evolution with temperature (the main cause of the UBR). Other models such as quasiparticle-phonon model (QPM), second RPA (SRPA), and extended RPA ($1p1h$ phonon or $2p2h$ phonon) are able to describe only the quantal width, but fail to include the thermal damping (thermal width), which is a unique feature and power of the PDM \cite{PDM1,PDM2}. Consequently, these models cannot fully reproduce the UBR. A recent finite-temperature RPA calculations has reproduced some low-energy $E$1 enhancements, which may arise from the redistribution of energy-weighted factors due to purely thermal effects in the Fermi-Dirac distribution \cite{Litvinova2013}. However, these effects do not reflect the true nature of UBR and fail to describe it accurately, as they not only mismatch experimental data but also lead to unrealistically high predicted UBR temperatures. Therefore, the principle that \textit{``Ideally, the theory has to include all correlations, continuum, and finite temperature"} \cite{Litvinova2013} is now fully integrated into our present approach. Moreover, our model does not encounter the computational limitations as in the shell model when applied to heavy nuclei.

Figure \ref{Fig2}(e) presents the ratio $R(\%)$ of the integrated strength of the RSF in the UBR region to that of the total RSF for X (X=GDR or GDR+UBR) calculated within the extended EP+PDM for all the considered UBR nuclei from light to heavy ones ($^{44,45}$Sc, $^{56,57}$Fe, $^{73,74,76}$Ge, $^{92,94,96,97}$Mo, $^{142,144,146,147}$Nd, and $^{151,153}$Sm). Here, $R$ is defined as $R= \int_{0}^{E} f^{UBR}(E_\gamma)dE_\gamma/\int_{0}^{E} f^X(E_\gamma)dE_\gamma$ with $E=30$MeV. This figure reveals that the UBR contributes approximately 5\% of the total RSF in light nuclei, while its contribution exponentially decreases as $A$ increases. This reflects an obvious trend that the UBR is generally present in light and medium nuclei, but becomes weaker or disappears with increasing $A$ and reappears in some heavy nuclei. A more comprehensive investigation with a larger statistical coverage is needed to fully validate this claim. Fitting the data in Fig. \ref{Fig2}(e) reveals  a global relation in the range of A=40-160, namely $R_{fit}=243.18\text{e}^{-A/11.45}+1.17\text{ln}A - 5.05$ with the coefficient of determination $\mathcal{R}^2=0.95$. For $E_{\rm UBR}$, the inset in Fig. \ref{Fig2}(e) shows global linear relations between $E_{\rm UBR}$ and $A$, namely $E_{\rm UBR} = a\times A + b$ with $a = -0.001273 \pm 0.002210, b = 0.957694 \pm 0.226534$ for even-$A$ systems and $a = -0.007367 \pm 0.002306, b = 1.288600 \pm 0.258252$ for odd-$A$ systems. The existence of these global trend suggests that both $E_{\rm UBR}$ and $\sigma_{\rm UBR}$ are not merely arbitrary values fitted to specific experimental data, but instead represent stable physical properties of the UBR that evolve predictably within the nuclear environment.

Last but not least, our study on the origin of the UBR also leads to a conclusion on the validity of the BAH. The temperature evolution of $f^{\rm UBR}$ for all the considered cases in Fig. \ref{Fig1} reveals that $f^{\rm UBR}$ is almost non-existent at $T=0$ and arises with strong temperature dependence at $T>0$. This result provides additional theoretical evidence for the invalidation of the BAH in the very low $E_\gamma$ region, in line with numerous previous studies \cite{Nabi,Farooq,Herrera,Sieja2023,Misch,HungPRL,Johnson,Angell,Isaak2013,Isaak2019,Netterdon,Wasilewska,Loens2012,Sieja2018,Isaak2018,Mercenne2024,Gorton2026}. It also offers a possible explanation for the deviations observed in BAH-based nuclear astrophysical calculations, where the UBR plays a crucial role \cite{Larsen2010,Spyrou2024}.\par

In conclusion, the present Letter has introduced a microscopic framework to describe the UBR based on the extended EP+PDM model. The UBR is interpreted in terms of a low-energy nuclear collective oscillation, termed the UBR phonon, which is coupled to $ph$, $pp$, and $hh$ configurations, following the same mechanism that causes the damping of the GDR phonon at finite temperatures. Our predictions for odd-odd, odd-$A$, and even-even UBR  nuclei, covering a wide range of $A$, are in good agreement with available RSF data, thus validating the proposed approach. Interestingly, the coupling strength of the UBR to $pp$ and $hh$ configurations is found to be about three times stronger than that of the GDR across all considered nuclei. In contrast, the contribution of coupling to $ph$ configurations is negligible in the description of the UBR. This feature reveals the thermodynamic nature of the UBR, as the $pp$ and $hh$ couplings are effective only at finite temperature. The predominantly $M$1 character of the UBR is reconfirmed in our calculations with only a small admixture of higher multipole excitations. Our findings also invalidate the BAH in the very low $E_\gamma$ region because the UBR emerges only at finite temperature. Finally, we obtain a global relation between the ratio of the UBR integrated strength to that of the total RSF and the nuclear mass number. The UBR description given in this Letter enhances the accuracy of nuclear reaction rate and Maxwellian cross-section calculations, with significant implications for nuclear astrophysics and other applications.

\acknowledgments
This work was funded by the National Foundation for Science and Technology Development (NAFOSTED) of Vietnam,
Grant No. 103.04-2025.22. L.T.P. would like to thank Dr. P. H. Khiem, T. T. Dzung, N. T. Huy, and T. V. Dong for their valuable discussions. N.N.A. gratefully acknowledges the support of Phenikaa University. 	

\end{document}